\documentstyle[11pt,newpasp,twoside,epsf]{article}
\def\fun{\hbox{\ erg cm$^{-2}$ s$^{-1}$} }

\def\Osix{O$\;${\footnotesize VI}$\;$}
\def\Ose{O$\;${\footnotesize VII}$\;$/O$\;${\footnotesize VIII}$\;$}
\def\cgs{erg~$\rm{s}^{-1}\rm{cm}^{-2}\rm{deg}^{-2}\;$}

\def\edcomment#1{\iffalse\marginpar{\raggedright\sl#1\/}\else\relax\fi}
\reversemarginpar

\begin{document}
\title{Studying the Evolving Universe with XMM-Newton and Chandra}
 \author{ G\"unther Hasinger }
\affil{ Max-Planck-Institut f\"ur extraterrestrische Physik \\
Postfach 1319, D--84541 Garching, Germany}

\begin{abstract}

Two X--ray observatories, the NASA observatory Chandra and the ESA mission XMM-Newton, provide powerful new diagnostics of the "hot universe". In this article I review recent X--ray observations of the evolving universe. First indications of the warm/hot intergalactic medium, tracing out the large scale structure of the universe, have been obtained lately in sensitive {\em Chandra} and {\em XMM-Newton} high resolution absorption line spectroscopy of bright blazars. High resolution X--ray spectroscopy and imaging also provides important new constraints on the physical condition of the cooling matter in the centers of clusters, requiring major modifications to the standard cooling flow models. One possibility is, that the supermassive black hole in the giant central galaxies significantly energizes the gas in the cluster.

{\em XMM-Newton} and {\em Chandra} low resolution spectroscopy detected significant Fe K$_\alpha$ absorption features in the spectrum of the ultraluminous, high redshift lensed broad absorption line QSO APM 08279+5255, yielding new insights in the outflow geometry and in particular indicate a supersolar Fe/O ratio. {\em Chandra} high resolution imaging spectroscopy of the nearby ultraluminous infrared galaxy and obscured QSO NGC 6240 for the first time gave evidence of two active supermassive black holes in the same galaxy, likely bound to coalesce in the course of the ongoing major merger in this galaxy.

Deep X--ray surveys have shown that the cosmic X--ray background (XRB) is largely due to the accretion onto supermassive black holes, integrated over the cosmic time. These surveys have resolved more than 80 \% of the 0.1-10 keV X--ray background into discrete sources. Optical spectroscopic identifications show that the sources producing the bulk of the X--ray background are a mixture of obscured (type--2) and unobscured (type--1) AGNs, as predicted by the XRB population synthesis models. A class of highly luminous type--2 AGN, so called QSO-2s, has been detected in the deepest {\em Chandra} and {\em XMM-Newton} surveys. The new {\em Chandra} AGN redshift distribution peaks at much lower redshifts (z $\approx$ 0.7) than that based on ROSAT data, indicating that the evolution of Seyfert galaxies occurs at significantly later cosmic time than that of QSOs. 
\end{abstract}

\section{Introduction -- Chandra and XMM-Newton}

The {\em Chandra X--ray Observatory} (CXO) -- previously called "Advanced X--ray Astrophysics Facility" (AXAF) -- was launched on July 23, 1999 and provided significant advances in observational capabilities: high-resolution ($\sim$ 0.5 arcsec) imaging and high resolution dispersive spectroscopy in the energy band 0.1--10 keV (see Weisskopf 1999). The 14m long, 4.5 ton spacecraft was placed in a highly elliptical 140000 km apogee, 10000 km perigee orbit. The X--ray telescope is a four-fold nested Wolter-I mirror system with Zerodur mirror shells of diameters between 0.63 and 1.2m and thickness between 16 and 24 mm. As previously demonstrated by the ROSAT mirrors, these shells can be made to very high optical quality and surface smoothness. The {\em Chandra} mirror system has perfected this performance with a point-spread function half-energy width (HEW) of $\sim$0.6 arcsec. 

{\em Chandra} has two focal plane instruments -- the High-Resolution Camera (HRC) and the Advanced CCD Imaging Spectrometer (ACIS), as well as two sets of objective transmission grating systems, which can be flipped into the optical path, the Low-Energy Transmission Gratings (LETG) and the High/Medium-Energy Transmission Gratings (HETG/METG). Each of the focal plane instruments has one detector for direct imaging and another one optimized for imaging of X--rays dispersed by the gratings. The HRC-I contains a 100mm microchannel plate and provides a position resolution of about 18 $\mu$m (0.4 arcsec). ACIS--I contains a 2x2 array of CCD detectors, each with 1024 x 1024 pixels of 24$\mu$m (0.5 arcsec). The transmission grating systems contain hundreds of grating facets mounted to 4 support rings. 

The {\em XMM-Newton Observatory}, a cornerstone mission of the ESA Horizon 2000 programme (Jansen et al., 2001), was launched on December 10, 1999 into a highly eccentric 48 hour orbit with a perigee of 7000 km and apogee of 114000 km. The 4 ton, 10m long spacecraft is the largest scientific satellite ever launched by ESA. A key technology component was the development of the large area X--ray mirror modules. Each of the three X--ray telescopes on board {\em XMM-Newton} consists of 58 Wolter I mirrors with shells replicated from superpolished gold coated mandrels using a nickel electroforming technique. The focal length is 7.5 m and the shell diameter and thickness are in the range 
306-700mm and 0.47-1.07mm, respectively. Two of the three mirror systems are equipped with Reflection Grating Spectrometers (RGS; den Herder et al. 2001) and an Optical Monitor (Mason et al. 2001) complements the front side of {\em XMM-Newton}.  

Apart from two CCD cameras for the RGS readout, the Focal Plane Assembly contains three imaging CCD cameras, provided by the European Photon Imaging Camera (EPIC) consortium. The two telescopes with reflecting grating spectrometer in the optical path are equipped with MOS type CCDs (Turner et al., 2001), the open beam telescope operates the novel pn-CCD as an imaging X--ray spectrometer (Str\"uder et al. 2001).

\begin{figure}[!ht]
\plotone{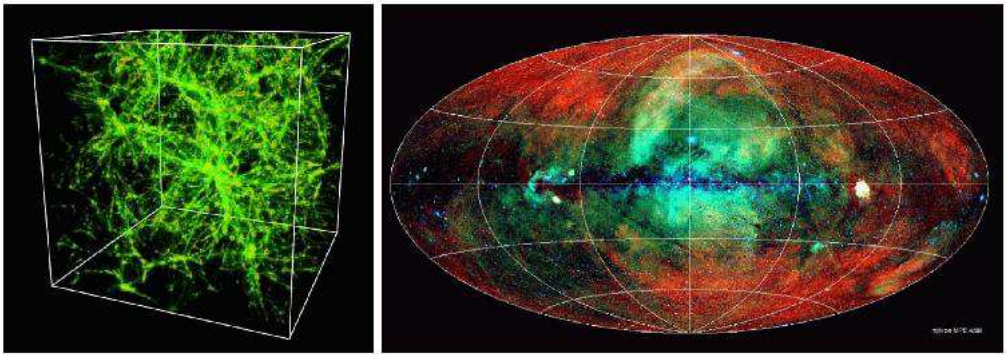}
\caption{\small \it Left: Spatial distribution of the warm/hot intergalactic medium (WHIM) with temperatures $10^{5-7}$K in a 100 Mpc box at $z=0$ from a hydrodynamical cold dark matter simulation with $256^3$ particles (Cen \& Ostriker, 1999) at z=0. Green, yellow and red refer to overdensity factors of 10, 100 and 1000, respectively.  Right: Diffuse X--ray background measured in the ROSAT All-Sky Survey (Freyberg, M., 2001, priv. comm.). Colours refer to different X--ray energies, red: 0.1-0.4 keV, green 0.5-0.9 keV, blue: 0.9-2.4 keV).}
\end{figure}

\section{Large-scale structure and the WHIM}

After the Big Bang, the dark matter in the universe coagulates into a pattern of filaments and voids - the "Cosmic web" of large scale structure. The baryons follow this structure formation, and at high redshift can be observed in the 
Lyman $\alpha$ forest. At low redshifts, however, the total census of
baryons in stars, interstellar medium and hot gas in clusters of galaxies 
is not enough to account for all baryons in the universe. According to cosmological hydrodynamical simulations (e.g. Cen \& Ostriker 1999), the 
gas in filaments should be shock-heated to temperatures of $10^5$-$10^7$K
in the course of the large-scale structure formation. In the local universe
almost half of the baryons should be in the form of a Warm/Hot Intergalactic
Medium (WHIM), observable mainly in X--rays. Figure 1a shows results of the 
Cen \& Ostriker hydrodynamical cold dark matter simulation. More recent
calculations (e.g. Dav\'{e} et al. 2000; Phillips et al. 2000) confirm this picture and improve on the predictions. Wu \& Xue (2001) present an estimate of the WHIM emission using the observed X--ray luminosity function and
luminosity-temperature relation for groups and clusters of galaxies and show
that the resulting 0.1-10 keV spectrum is consistent with the most recent 
hydrodynamical models as well as the current upper limits on the unresolved fraction of the X--ray background (XRB; see below). 

The ROSAT All-Sky Survey background map does not reveal the expected filamentary structure of the WHIM emission (see Fig. 1b). The X--ray background is dominated by local galactic structure and the cosmic X--ray background, which is largely resolved into discrete sources (see below). The low density and temperature of the warm/hot gas imply a low emissivity and make it difficult to detect it in emission. An upper bound of $1.1 \times 10^{-12}$ \cgs  in the 0.5 to 2 keV band has been derived by Briel \& Henry (1995) for filaments of hot gas between clusters. The WHIM may have been indirectly detected in emission in the autocorrelation function of the XRB (Sliwa, Soltan \& Freyberg, 2001) and in the cross-correlation function between the XRB and clusters of galaxies (Soltan et al. 1996). Recently, however, warm and hot gas possibly surrounding our galaxy has been detected in absorption as \Osix absorption features in FUSE ultraviolet spectra of quasars (Tripp et al., 2001) and as 
\Ose absorption features in Chandra/XMM-Newton X--ray quasar spectra (Nicastro 
et al., 2002; Fang et al., 2002; Paerels et al., 2003), confirming the basic predictions of the CDM hydrodynamic models. 

\begin{figure}[!ht]
\plotone{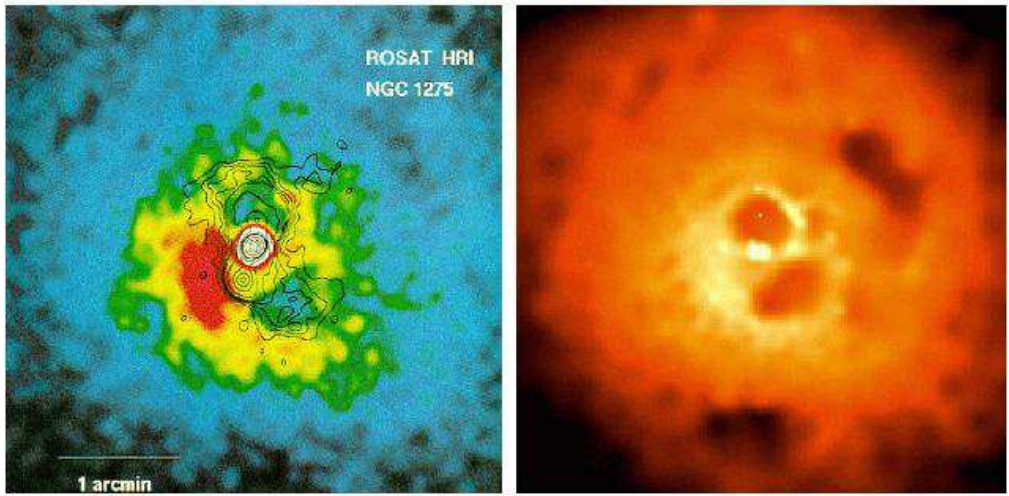}
\caption{\small \it X--ray images of the detailed structure of the intergalactic medium around the galaxy NGC 1275 in the center of the Perseus cluster. Radio contours are superposed on the {\em ROSAT HRI} image (left; B\"ohringer et al., 1993). The structure is seen at higher detail and contrast in the {\em Chandra ACIS} image (right; Fabian et al., 2000).}
\end{figure}

\section{Complex centers of clusters of galaxies}

Clusters of galaxies represent the largest virialized structures in the Universe. They form at the nodes of the "Cosmic web" and contain 
$10^{13-15} M_{\odot}$ of dark matter and up to several thousand galaxies, typically with a giant galaxy at the center. Apart from the dark matter, the galaxies swim in a sea of hot intergalactic medium of a temperature of several million K - the extension of the WHIM discussed in the previous section (see Fig. 1a). The X--ray emissivity of hot gas is proportional to the square of the gas density and therefore, unlike in the case of the filaments, the diffuse X--ray emission from clusters is rather bright. It was detected with the first X--ray satellites UHURU, Ariel V and HEAO-1 in the seventies and has been studied thereafter with ever higher precision with the Einstein, ROSAT and ASCA observatories.

Since the gas density is highest in the cluster center, the emissivity is highest there and correspondingly the cooling time of the gas in the center of large clusters is shorter than the age of the Universe. According to the standard "cooling flow" model (see Fabian 1994 for a review), the gas in the central area should cool in about a billion years and should form stars at a rate of up to $100~M{_\odot}$/yr. The cooling flow model was, however, debated because neither the cool gas itself nor the young stars formed could be observed in the cluster centers.

New observations with XMM-Newton and Chandra have now provided completely new and unexpected results from the central regions of clusters. The large dispersion and high sensitivity of the Reflection Grating Spectrometer on XMM-Newton make it the ideal instrument for observations of the cooling flows. For the first time it is possible to detect emission lines of individual ions, in particular of the Fe L shell series which, in the case of a cooling flow should dominate the X--ray emission for temperatures below $\sim$2 keV. The 
high-resolution XMM grating spectra of a number of nearby, bright clusters, however, do not show any of the emission lines expected from temperatures below 
1--2 keV. Lines from somewhat hotter gas are visible, indicating that gas is cooling down to about 2-3 keV from on average significantly higher temperatures, but the amount of cooler gas is at least a factor of ten lower than expected from simple cooling flow models (Peterson et al., 2001; Tamura et al., 2001). XMM-Newton EPIC CCD observations of M87 in the center of the Virgo cluster, although on one hand with significantly worse energy resolution, on the other hand much better photon statistics and angular resolution than the dispersed X--ray spectra, confirm this result and also rule out any excess absorption by cold gas in the centers of the Virgo and Perseus cluster (B\"ohringer et al., 2002). 

High angular resolution Chandra images of some nearby clusters, following on earlier ROSAT results, reveal a remarkable complexity of structure in the centers of nearby clusters and an entanglement between the X--ray gas and the radio jets of the core galaxies (Fabian et al., 2000; McNamara et al, 2001). Fig. 2a shows the ROSAT HRI image of NGC1275, the radio galaxy at the center of the Perseus cluster with radio contours overlaid (B\"ohringer et al., 1993). There is a one to one correspondence between the radio lobes and darker regions in the central hot X--ray emitting gas. The Chandra observation of the same galaxy (Fig. 2b) from Fabian et al. (2000), confirms this structure in much higher detail and contrast. The radio sources are apparently pushing away the hot gas and produce cavities filled with radio emission and probably pressure supported by magnetic fields and cosmic rays. In addition, radio-faint, ``ghost'' cavities are seen in some clusters, which may be relics of earlier AGN activity floating in the intracluster medium. 

Various possibilities to reconcile the new observations with variants of the 
cooling flow models have been proposed, like e.g. heating, mixing, differential absorption and inhomogeneous metallicity (see Fabian et al., 2001). B\"ohringer et al. (2002) suggest a connection between the absence of cool gas, the presence of powerful radio jets from the supermassive black hole and the existence of the buoyant, magnetically supported bubbles in nearby clusters. If the hot cluster gas -- which clearly cools towards the center -- does not cool down below a certain temperature, one possible solution is, that it must be constantly heated. B\"ohringer et al. argue that the energy liberated by accretion onto the black hole in the central galaxy, which grows to between several million and a billion solar masses in the lifetime of the Universe, is by far dominating the cooling energy of the central cluster gas. In the case of M87, according to the model by Churazov et al. (2001), we seem to observe the action of the radio jet on the cluster gas directly. One problem with this picture is the necessary fine tuning between cooling and heating in the cluster center, which could be provided by a self-regulating mechanism of accretion occurring at the Bondi rate. As pointed out by Fabian et al. (2000) it is, however, difficult to equilibrate the necessary heating steadily over time and over the whole cluster volume. Possible ways out may be the AGN activity of all black holes in all the galaxies distributed throughout the cluster or large-scale magnetic effects that distribute the heat. Nevertheless, it is likely that the central black holes, which we now know to exist in most large galaxies, play an important role in the evolution of the gas around the galaxies.

\section{Iron at high redshift}

Broad absorption Line (BAL) Quasi-stellar Objects (QSOs) are a key to understand the geometry and physical state of the medium in the immediate vicinity of accreting supermassive black holes. A new unified model (Elvis 2000) indicates that a significant fraction of the matter accreted into the region of the compact object is flowing out again. On either side of the accretion disk it should form a thin funnel-shaped shell, in which the outflowing gas is ionised and accelerated to velocities of 0.05-0.1c by the powerful radiation force of the central object. If the observer's inclination is favourable, this flow intercepts the line of sight with a large column density and produces the blue-shifted broad UV absorption line features observed in about 10\% of all QSOs. However, the UV/optical spectra sample only a minor fraction of the total column density of the flow, which is predicted to be highly ionised so that it mainly absorbs X--rays.

\begin{figure}[!ht]
\plotone{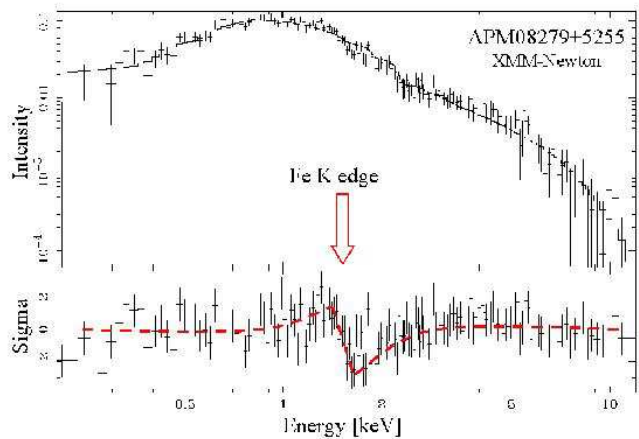}
\caption{\small \it X--ray spectrum of APM 08279+5255 taken with the XMM--Newton pn--CCD camera, fit with a simple power law model absorbed by neutral gas in our own galaxy as well as associated with the source. This fit is 
statistically not acceptable; the residuals show systematic deviations which 
are interpreted as an ionised Fe-K edge (Hasinger, Schartel \& Komossa, 2002).}
\end{figure}

APM 08279+5255 is an exceptionally luminous Broad Absorption Line (BAL) QSO at redshift z=3.91, originally identified serendipitously in a survey for Galactic halo carbon stars (Irwin et al., 1998). The source is an ultraluminous IRAS galaxy with a far-infrared luminosity of $> 5 \times 10^{15}~L_{\odot}$. The object is strongly lensed (Ledoux et al., 1998), with a magnification factor of 50-100, but even taking this magnification factor into account, the object is still among the most luminous in the Universe. The optical spectrum of APM 08279+5255 shows a broad absorption trough (BAL) on the blue side of Ly$\alpha$, indicating a heavily structured and highly ionised gas stream, outflowing with velocities up to 12000 km/s (Srianand \& Petitjean, 2000). 

APM 08279+5255 was observed with {\it XMM--Newton} in October 2001 and April 2002 with exposure times of 16 and 100 ksec, respectively. The high redshift and high luminosity of this object, combined with the superb sensitivity of the {\it XMM--Newton} instruments allowed the detailed X--ray diagnostic of a BAL flow with unprecedented precision (Hasinger, Schartel \& Komossa 2002). The X--ray spectra show systematic residuals around the rest frame energy of iron K$\alpha$ (see Fig. 3), which are interpreted as an absorption edge of significantly ionised Fe-K$\alpha$ (Fe XV - XXVIII) from a high column density absorber ($N_H>10^{23}~cm^{-2}$), very likely associated with the highly ionised BAL flow observed in the UV spectrum of the source. These results therefore confirm a basic prediction of the phenomenological geometry models for the BAL outflow (Elvis 2000) and can constrain the size of the absorbing region. 

The low energy absorption in the spectra spectra, which is due to the Fe-L edge and the K-edges of lower-Z elements can also constrain the Fe abundance. Surprisingly, the Fe/O abundance of the absorbing material is significantly higher than solar (Fe/O = 2--5), giving interesting constraints on the gas enrichment history in the early Universe. The outflowing material must have already been processed in a starburst environment. Because the iron enrichment depends mostly on the lifetime of SN Ia precursors, it takes $\sim$1Gyr until Fe/O reaches solar values. Fe measurements in high-$z$ objects, like APM 08279+5255, are therefore of profound relevance for understanding the early star formation history of the universe, and provide important constraints on cosmological models. Given the strong indications of a supersolar Fe/O abundance, APM 08279+5255 is beginning to constrain cosmological models, favouring those which predict larger galaxy ages at a given $z$. A recent Chandra observation of APM 08279+5255 confirms the presence of Fe absorption features in the energy range observed by XMM-Newton (Chartas et al., 2002). However, the detailed shape of the Chandra spectrum is different from that observed by XMM-Newton, indicating significant variability in the BAL flow. This object is therefore a candidate for repeated observations by both XMM-Newton and Chandra.

\section{A binary AGN in NGC 6240}

NGC6240 is one of the nearest members of the class of Ultraluminous infrared galaxies (ULIRGs), which are outstanding due to their huge infrared luminosity output, predominantly powered by starburst processes and/or hidden active galactic nuclei (AGN). The galaxy shows conspicuous loops and tails, and two optical, IR and radio nuclei separated by $\sim 1.8"$. NGC6240 is in the middle of a major merger and is expected to finally form an elliptical galaxy (see Komossa et al., 2003 and references therein).

NGC6240 was observed in July 2001 with the Chandra ACIS-S instrument with an effective exposure time of 37 ks. Figure 4 shows the Chandra X--ray image of NGC6240 in the 0.5-8 keV band with an inset showing only the 5-8 keV band. 
At X--ray energies below 2.5 keV, extended, loop-like emission and several knots appear, which are well correlated with the HST $H_\alpha$ images (Gerssen et al. 2001) and clearly indicates a starburst-driven superwind activity. The rich soft X--ray structure changes as a function of energy and may indicate different stages of superwind activity. 

\begin{figure}[!ht]
\plotone{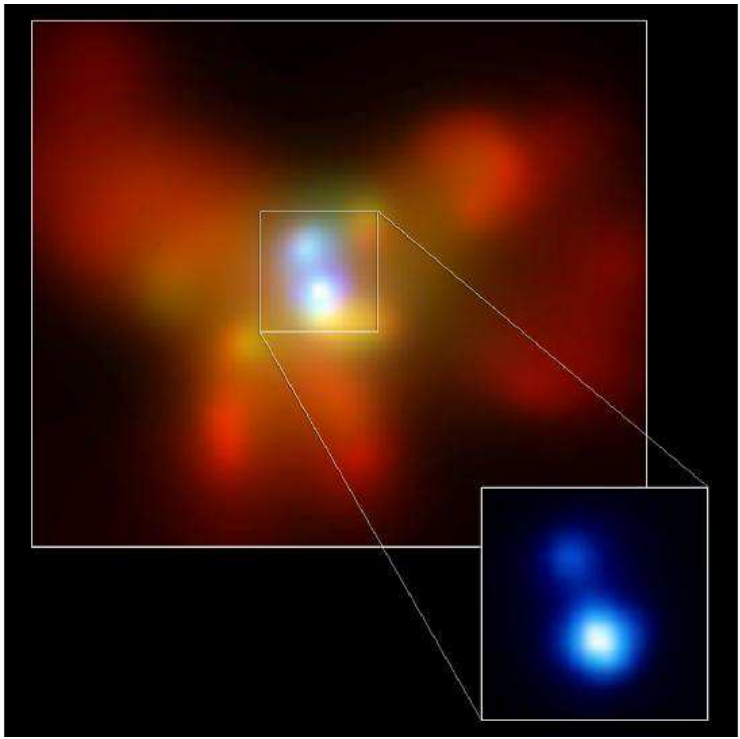}
\caption{\small \it 
The Chandra image of NGC 6240, a butterfly-shaped galaxy that is the product of the collision of two smaller galaxies, revealed that the central region of the galaxy (inset; 5-8 keV) contains not one, but two active giant black holes  
(Komossa et al., 2003).}
\end{figure}

The hard X--ray image (5-8 keV) is dominated by emission from two compact sources
(inset Fig. 4), spatially coincident within the errors with the IR position of the two nuclei of NGC6240. The AGN character of both nuclei is revealed by the detection of absorbed hard, luminous X--ray emission and two strong neutral Fe K$_\alpha$ lines around 6.4 keV from fluorescence in cold material
illuminated by a hard continuum. For the first time it is possible to disentangle the contribution to the hard X--ray luminosity from the southern and northern nucleus. The observed 0.1-10 keV X--ray luminosity of the two nuclei is $L_{x,S} = 1.9 \cdot 10^{42}$ erg/s, $L_{x,N} = 0.7 \cdot 10^{42}$ erg/s, respectively, with an observed absorbing column density around $10^{22}$ cm$^{-2}$, which is likely due to scattering, because the intrinsic, much more luminous and more absorbed AGN emission of NGC6240 only shows up above $\sim$9-10 keV (Vignati et al. 1999). The strength of the neutral Fe K$_\alpha$ line and the flatness of the spectra in both nuclei suggests a scattering geometry for both AGN. 

Ultimately, the binary AGN of NGC6240 will coalesce to form one nucleus. The final merging of the supermassive black holes is expected to produce a strong gravitational wave signal. In fact, such events are expected to generate the clearest signals detectable with the gravitational wave detector LISA that will be placed in Earth orbit in the near future (e.g. Vitale 2002).

\section{Resolving the X--ray background}

Black holes in the centers of galaxies are also at the core of another important observational advance with {\em Chandra} and {\em XMM-Newton}. Deep X--ray surveys indicate that the cosmic X--ray background (XRB) is largely 
due to accretion onto supermassive black holes, integrated over cosmic time. In 
the soft (0.5-2 keV) band more than 90\% of the XRB flux has been resolved using 
1.4 Msec observations with {\em ROSAT} (Hasinger et al., 1998) and recently 1--2 Msec {\em Chandra} observations (Brandt et al., 2001; Rosati et al., 2002; Giacconi et al., 2002) 
and 100 ksec observations with {\em XMM-Newton} (Hasinger et al., 2001). In the harder (2-10 keV) band a similar fraction of the background has been resolved with the above {\em Chandra} and {\em XMM-Newton} surveys, reaching source densities of about 4000 deg$^{-2}$ (see Fig. 5). 

The X--ray observations have so far been about consistent with population 
synthesis models based on unified AGN schemes (Comastri et al., 1995; Gilli, Salvati \& Hasinger, 2001), which explain the hard spectrum of the X--ray background by a mixture of absorbed and unabsorbed AGN, folded with the corresponding luminosity function and its cosmological evolution. According to these models, most AGN spectra are heavily absorbed and about 80\% of the light produced by accretion will be absorbed by gas and dust (Fabian et al., 1998). In particular they require a substantial contribution of high-luminosity obscured X--ray sources (type--2 QSOs), which previously have only been scarcely detected.

\begin{figure}[!ht]
\plotone{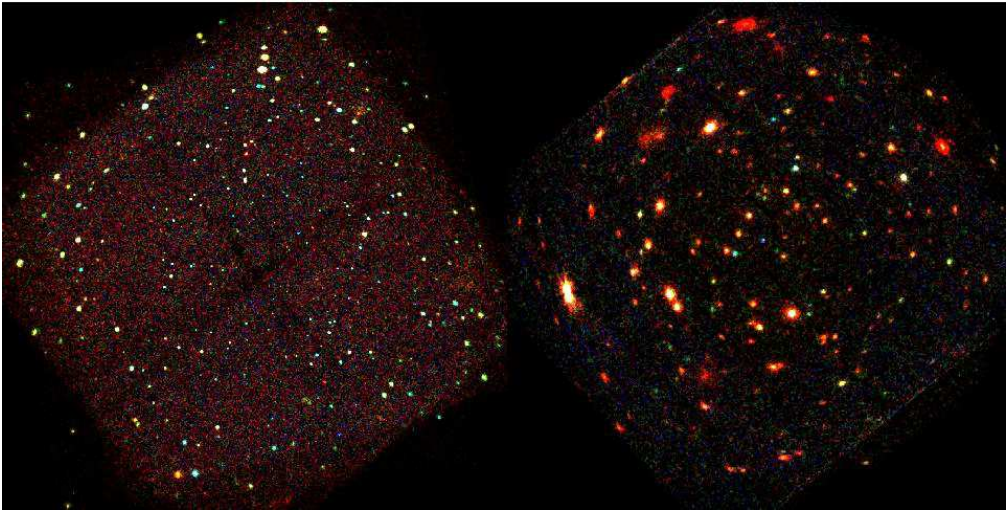}
\caption{\small \it Deep X--ray surveys: {\em Chandra} ACIS--I image of the {\em Chandra Deep Field South} (left, Rosati et al., 2002) and the {\em XMM-Newton} PN and MOS images of the {\em Lockman Hole} field (right, Hasinger et al., 2001). The field sizes are about 20$\times$20 arcmin and 30$\times$30 arcmin, respectively.} 
\end{figure}

Optical follow-up programs with 8-10m telescopes have been completed for the ROSAT deep surveys and find predominantly AGN counterparts of the faint X--ray source population (e.g. Lehmann et al., 2001) mainly X--ray and optically unobscured AGN (type--1 Seyferts and QSOs) and a smaller fraction of obscured AGN (type--2 Seyferts). Optical identifications for the deepest {\em Chandra} and {\em XMM-Newton} fields are now approaching a completeness of 60-80\% and find the predicted mixture of obscured and unobscured AGN with an increasing fraction of obscuration towards fainter fluxes (Barger et al., 2002; Szokoly et al., 2003; see below). Interestingly, first examples of the long--sought class of type--2 QSO have been detected in deep {\em Chandra} fields (Norman et al., 2002; Stern et al., 2002a) and in the {\em XMM-Newton} Deep survey in the {\em Lockman Hole} field (Hasinger et al., 2001b). 

X--ray diagnostics using luminosity and hardness ratio of the individual objects can be used in addition to the optical spectroscopy to classify the sources identified so far in the different fields. Table 1 gives an update of the current identification content in some of the major deep fields.

\begin{table}
\caption[]{{\small \it Chandra} and {\it XMM-Newton} Deep Survey Identifications}
\begin{center}
\begin{tabular}{lrrrrl}
\hline\hline
Field   &        typ1 & typ2 & gal & tot & Reference \\
\hline
HDF--N     &  57     &  60    & 55  &  172  &Barger et al., 2002\\
CDFS          &  44     &  58    & 33  &  135  &Szokoly et al., 2003\\
CDFS$^a$      &  80     & 103    & 71  &  254  &Wolf; Mainieri, priv. comm.\\
Lockman Hole  &  41     &  26    &  7  &   74  &Lehmann et al. 2001\&2\\
Abell 370     &   9     &   5    &  6  &   20  &Barger et al., 2001b\\
13hr$^b$      &   5     &   7    &  1  &   20  &Barger et al., 2001a\\
Lynx 3.A      &   7     &   8    &  2  &   19  &Stern et al., 2002a\&b\\
\hline
\end{tabular}
\leftline{$^{\rm a}$  including photometric redshifts} \\
\leftline{$^{\rm b}$  only 2-7 keV band detections considered}
\label{tab:surv}
\end{center}
\end{table}

After having understood the basic contributions to the X--ray background, the 
general interest is now focussing on understanding the physical nature of these 
sources, the cosmological evolution of their properties, and their role in 
models of galaxy evolution. The luminosity function of X--ray selected AGN based on the ROSAT data shows strong cosmological density evolution at redshifts up to 2, which goes hand in hand with the cosmic star formation history (Miyaji et al., 2000; 2001), but still lacks statistical significance at high redshifts and low luminosities. The new {\em Chandra} are now providing additional constraints on the high redshift evolution of QSO and the behaviour of lower luminosity Seyfert galaxies at redshifts up to $\sim$2.

\section{The new Chandra redshift distribution}

Most of the above samples have a spectroscopic completeness of about 
60\%, which is mainly caused by the fact that about 40\% of the 
counterparts are optically too faint to obtain reliable spectra. 
The completeness can be improved to about 80 \%, when photometric 
redshifts are included. This completeness therefore allows to compare the observed redshift distribution with predictions from X--ray background population synthesis models (Gilli, Salvati \& Hasinger 2001), based on the AGN X--ray luminosity function and its evolution as determined from the ROSAT surveys (Miyaji et al., 2000), which, due to the saturation of the QSO evolution  predict a maximum at redshifts around z=1.5. Figure 6 shows two predictions of the redshift distribution from the Gilli et al. model for a flux limit of 
$2.3 \times 10^{-16}~\fun$ in the 0.5-2 keV band with different assumptions for the high-redshift evolution of the QSO space density. The two models have been normalized at the peak of the distribution.

\begin{figure}[!ht]
\plotone{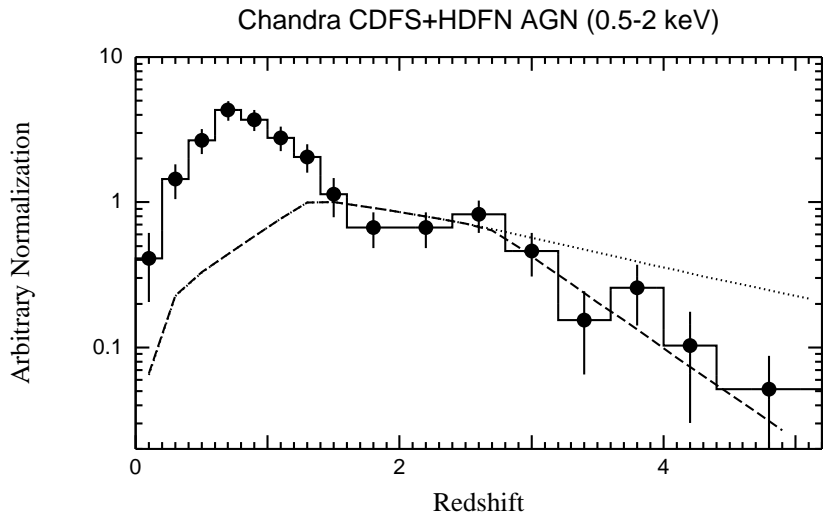}
\caption{\small \it Redshift distribution of 243 AGN selected
in the 0.5-2 keV band from the inner 10 arcmin radius of the Chandra CDFS and HDFN survey samples (solid circles and histogram), compared to model predictions from population synthesis models (Gilli et al., 2001). The dashed line shows the prediction for a model, where the comoving space density of high-redshift QSO follows the decline above z=2.7 observed in optical samples (Schmidt, Schneider \& Gunn, 1995; Fan et al., 2001). The dotted line shows a prediction with a constant space density for $z>1.5$. The two model curves have been normalized to their peak at z=1, while the observed distribution has been normalized 
to roughly fit the models in the redshift range 1.5--2.5}
\end{figure}

The actually observed redshift distribution of AGN selected from the HDF--N and CDFS Chandra deep survey samples (see Table 1) at off-axis angles below 10 arcmin and in the 0.5-2 keV band has been arbitrarily normalized to roughly fit the population synthesis models in the redshift range 1.5 - 2.5 keV an shown in Fig. 6 as histogram and data points. In the redshift range below 1.5 it is radically different from the prediction, with a peak at a redshift at z$\approx$0.7. This low redshift peak is dominated by Seyfert galaxies with X--ray luminosities in the range $L_X = 10^{42-44}$ erg/s. Since the peak in the observed redshift distribution is expected at the redshift, where the strong positive evolution of AGN terminates, we can conclude that the evolution of Seyfert galaxies is significantly different from that of QSOs, with their evolution saturating around a redshift of 0.7, compared to the much earlier evolution of QSOs which saturates at z$\approx$1.5. The statistics of the two samples is now sufficient to rule out the constant space density model at redshifts above 3, clearly indicating a decline of the X--ray selected QSO population at high redshift consistent with the optical findings. However, the statistical errors and the likely spectroscopic incompleteness still preclude a more accurate determination.

\end{document}